\documentstyle[twocolumn,prc,aps,epsfig]{revtex}

\input epsf
\newcommand{\be}{\begin{eqnarray}}
\newcommand{\ee}{\end{eqnarray}}

 \newcommand{\gsim}{\mathrel{\hbox{\rlap{\lower.55ex \hbox {$\sim$}}
                   \kern-.3em \raise.4ex \hbox{$>$}}}}
\newcommand{\lsim}{\mathrel{\hbox{\rlap{\lower.55ex \hbox {$\sim$}}
                   \kern-.3em \raise.4ex \hbox{$<$}}}}

\def\roughly#1{\mathrel{\raise.3ex\hbox{$#1$\kern-.75em%
\lower1ex\hbox{$\sim$}}}}
\def\lsim{\roughly<}
\def\gsim{\roughly>}

\renewcommand{\thefootnote}{\arabic{footnote}}
\begin{document}

\twocolumn[\hsize\textwidth\columnwidth\hsize\csname @twocolumnfalse\endcsname

\title{ Electric Flux Tube in Magnetic Plasma}
\author {Jinfeng Liao and Edward Shuryak}
\address {Department of Physics and Astronomy, State University of New York,
Stony Brook, NY 11794}
\date{\today}
\maketitle
\begin{abstract}
In this paper we study a methodical problem related to the
magnetic scenario recently suggested and initiated by the authors
\cite{Liao_ES_mono} to understand the strongly coupled quark-gluon
plasma (sQGP): the electric flux tube in monopole plasma. A
macroscopic approach, interpolating between Bose condensed (dual
superconductor) and classical gas medium is developed first. Then
we work out a microscopic approach based on detailed quantum
mechanical calculation of the monopole scattering on electric flux
tube, evaluating induced currents for all partial waves. As
expected, the flux tube looses its stability when particles can
penetrate it: we make this condition precise by calculating the
critical value for the product of the flux tube size times the
particle momentum, above which the flux tube dissolves. Lattice
static potentials indicate that flux tubes seem to dissolve at
$T>T_{dissolution} \approx 1.3 T_c$. Using our criterion one gets
an estimate of the magnetic density $n\approx 4.4 \sim 6.6
fm^{-3}$ at this temperature.
\end{abstract}
\vspace{0.6in} ]
\newpage

\section{Introduction}

\subsection{Electric-magnetic duality and monopoles}

Studies of magnetically charged objects resurface in theoretical
literature regularly since Maxwell's time, as soon as some new
developments call for their application in different fields. Three
of these are especially important: (i) the celebrated Dirac
quantization condition
 \cite{Dirac}, (ii) the idea put forward by 't Hooft and Mandelstam
 \cite{'t Hooft-Mandelstam} to view
 the QCD confining vacuum as a ``dual
superconductor'', and (iii) Seiberg and Witten \cite{SeiWit}
studies of  ${\cal N}=2$ SUSY gauge theories, identifying
magnetically charged degrees of freedom and the usefulness as well
as necessity of jumping from ``electric'' to ``magnetic'' language
whenever appropriate. There are excellent reviews covering
historic development of these ideas, see
\cite{Goldhaber,Milton,Harvey_review}, and (for more recent
progress) \cite{Weinberg_review,Shifman_review}, and also the
excellent book \cite{Shnir_book} by Shnir.

In the context of QCD phase diagram (finite temperature $T$ -
 baryonic chemical potential $\mu$) the issue of electric-magnetic
duality were discussed in our previous paper \cite{Liao_ES_mono}
(to be referred below as LS1). Brief summary can be made by noticing
that since electric and magnetic objects generally repel each
other, their simultaneous presence results in dominance of one
over the other. In particular: (i) at high-T the
Quark-Gluon Plasma (QGP) is
dominated by electric objects (gluons
and quarks); (ii) at high-$\mu$ and low $T$
QCD  is in  a color superconductor phase,
dominated by electrically charged diquarks  and confining magnetic
objects; while (iii) the vacuum is presumably
a dual (magnetic) superconductor
confining electric objects.

The most important point made in
LS1 is an
introduction of new ``electric-magnetic equilibrium line'',
 on which both couplings (electric and magnetic), densities, screening
 lengths etc are about equal, and thus the volume is somehow shared
equally.
 The Dirac condition then demands that at the equilibrium
line both electric and magnetic ``fine structure
constants''\footnote{Note in this work we change units from LS1
and use couplings normalized as standardly done in gauge theories,
with $\alpha=g^2/4\pi$ in the Coulomb law, i.e. Heaviside-Lorentz
units. } are \be {e^2 \over 4\pi} ={g^2 \over 4\pi}=1 \ee
Furthermore, as pointed out by 't Hooft, in nearly all phenomena
the relevant effective coupling is not $e,g$ themselves
but $\lambda_e=e^2 N_c$
or $\lambda_g=g^2 N_c$, further increasing with
the number of colors $N_c$. Thus one naturally is lead to the
conclusion that the
QCD plasma at e/m equilibrium conditions must be  strongly
coupled. In LS1 this was proposed to be a possible explanation of
why experiments at RHIC have indeed found a strongly coupled QGP,
see \cite{Shuryak_03,discovery_workshop,Shuryak_06,Shuryak_07} for
discussion of data and other theoretical ideas.

The QCD vacuum and deconfining transition was attempted to be
described by an ensemble of self-dual dyons in
\cite{Ilgenfritz_dyon}\cite{Diakonov_dyon}. The presence of
magnetic monopoles, both below and above the deconfinement phase
transition was studied in vast literature on lattice gauge
theories, see e.g. \cite{lattice_mono}. Our previous paper LS1 has
also studied interrelation of electric and
 magnetic quasiparticles numerically, but using much simpler
classical tool -- the Molecular Dynamics (MD). While lattice study
uses Euclidean time formalism and thus is restricted to
thermodynamical observables only: we on the other hand were able
to do real-time simulation and calculate such kinetic properties
as viscosity and diffusion constant, see \cite{Liao_ES_mono} for
more details.

\subsection{Electric Flux Tube Formation in Magnetic Plasma}

After this brief introduction, let us turn to the subject of the
present paper, the flux tubes. In the usual electric
superconductor the corresponding solution of Ginzburg-Landau
equations was first found by Abrikosov \cite{Abrikosov}, which was
later revived in field theory \cite{Nielsen_Olesen} and became
known as Abrikosov-Nielsen-Olesen (ANO) vertex. In the ``dual
superconductor'' picture of the QCD vacuum properties of the QCD
confining string and the resulting heavy quark potentials have
been discussed extensively: see e.g. reviews by M. Baker
\cite{Baker:1991bc}, and more recetnly by G. Ripka\cite{Ripka}
(with exhaustive list for further references).

Lattice studies (e.g. \cite{Balistrings})
provided substantial support to these works. Flux tube behavior at
finite $T$ was also extensively discussed: in particular Polyakov
\cite{Polyakov} has shown how exponential growth of flux tube
entropy leads to vanishing of the effective tension in free energy
$F(r,T)$ and Hagedorn-like phase transition.
This scenario would predict gradual deconfinement with the string
tension vanishing at $T_c$: in fact for $N_c>2$ it jumps to
zero.
 Deconfinement
transition for various number of colors $N_c$ was studies in detail:
see e.g. \cite{Bringoltz:2005kx} where $N_c$ up to 12 was studied.
Working with metastable ``overheated'' confined phase it
 was  found that the Hagedorn-like transition (at which the string tension of the free energy
vanishes
$\sigma(T)\rightarrow 0$)  can be approximately
located
at a universal ($N_c$ independent)
$T_H/T_c=1.116(9)$.

 Heating usual superconductors above the critical temperature
destroys not only the condensate but also
Cooper pairs themselves. Although normal (metallic) phase is
a plasma of electric objects (electrons), but their characteristic
momenta $p\sim p_F$ are orders of magnitude larger than
momenta of Cooper pairs,  thus there is no analog
of Abrikosov vortexes in the normal phase.
 This does not happen because presence of
a quantum condensate is that necessary for flux tube's existence:
 a counterexample can be provided e.g. by quite spectacular
magnetic flux tubes in solar plasma\footnote{ They have very large
fluxes and sizes, and thus a macroscopic theory --
magnetohydrodynamics -- can be
used for their description, which
 unfortunately it is not applicable in our case, for
microscopically small tubes.}. Whether charges are Bose-condensed
 or not, their scattering on a
flux tube may provide a  pressure which may lead to its
stabilization. It is just a matter of certain quantitative condition
for tube stabilization being met.

 The questions to be addressed in this work
 is whether QGP is like an electric plasma in a metal, without
magnetic flux tubes, or like other plasmas which have them? What
exactly are the necessary conditions for a flux tube formation?
Below we will ignore electric quasiparticles which would induce
screening/termination of electric flux lines\cite{Liao_ES_polymer}
and consider purely magnetic plasma. We will perform
quantum-mechanical study of monopole scattering on the tube and
examine their back-reaction to the tube field through the
associated magnetic current. This will answer these questions.

But before we do so, let us explain few important issues {\it
classically}, related to the very essence of the electric-magnetic
competition, i.e. ``expulsion'' of sub-dominant component into
flux tubes and their stabilization. A full quantum mechanical
calculation will be presented in Sections III-V.

We first start with an electric charge $e$ being placed within a
free gas of monopoles with mass $M$ and charge $\pm g$. The
monopole gas should be neutral, i.e. with equal number of positive
and negative charges. We emphasize in advance that monopoles with
either signs have the same effect: this will be seen in the
appearance of $g^2$ rather than $\pm g$ in the final results.

\begin{figure}[t]
  \epsfig{file=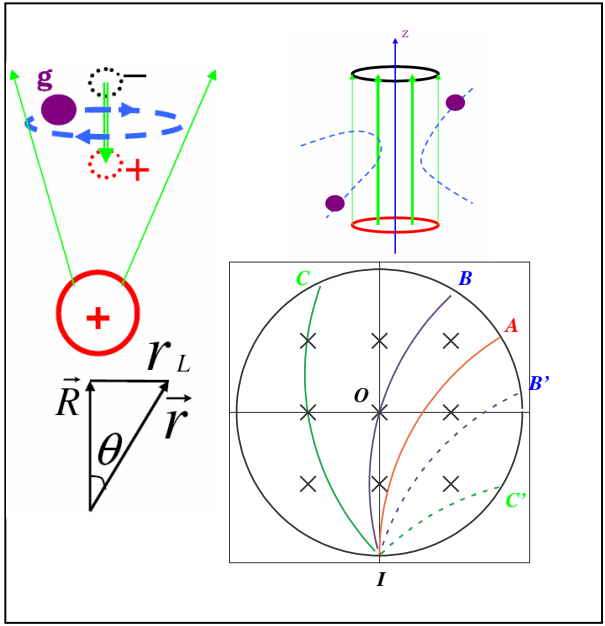,width=8.cm}
  \vspace{0.15in}
\caption{(color online) Schematic demonstration of anti-screening
effect and flux tube formation in magnetic medium, see text for
detailed discussions. } \label{fig_demonstration}
\end{figure}

At a distance $\vec R$ from the charge (see
Fig.\ref{fig_demonstration} left), the (unmodified) electric field
$\vec E_{\vec R}=\frac{e}{4\pi R^2}\hat{R}$ will stir the magnetic
monopoles into Larmor motion with radius $r_L$. As Poincare has
shown\cite{Goldhaber,Milton,Shnir_book} a century ago, the radius
shrinks near the charge, restricting the motion to a cone --- a
small patch of the whole space solid angle. The cone angle is
determined by (with $v_t$ the monopole velocity transverse to
$\hat{r}$)
\begin{equation}
\cot\theta = \frac{(ge)/4\pi c}{M v_t r}
\end{equation}
The numerator is precisely the field angular momentum of a
charge-monopole pair $L_{EM}=(ge)/4\pi c$ as first computed by J.
J. Thompson in 1896\cite{Goldhaber,Milton,Shnir_book}, while the
denominator is the monopole's kinetic angular momentum $L_v=M v_t
r$ with respect to the origin. The above formula, rewritten as
$\cot\theta=L_{EM}/L_v$, reflects the interplay between angular
momenta of the electromagnetic field and of the particle motion.
Though superficially $L_v$ is defined through $v_t$ and $r$, it is
actually a conserved quantity uniquely related to the cone angle
$\theta$, see \cite{Shnir_book} for detailed discussion.

In turn, these monopoles form loops of magnetic current $ g n_L
L_v/Mr $ ($n_L$ their density) on the cone. The direction of the
current explains the sign of induced electric dipole\footnote{Note
that although monopoles with $\pm g$ rotate in opposite
directions, they produce currents of the same sign, so it is not
necessary to distinguish them.}. Using dual Maxwell's equation
$\vec \bigtriangledown \times \vec E = -\frac{1}{c} \vec J_M$, one
finds that such electric dipole is opposite to induced dipoles in
dielectric, so in this sense it is an {\em anti}-screening effect.
The charge repels such a dipole: thus monopole will fly away from
the charge.

To make this statement quantitative, let's calculate the curl of
magnetic current around $\vec R$. To do that we need to require
that the Larmor circle to be fairly small, for two important
reasons: (i) if it is not small then one has to take into account
the variation of electric field strength which will warp the
circle; (ii) a small Larmor radius enables one to approximate the
$\vec \bigtriangledown \times J_M$ by integrating $\vec J_M$ along
the circumference and dividing it by the area of the circle. Small
Larmor radius $r_L$ means small angle $\theta$, i.e. $L_v <<
L_{EM}$. Density of monopoles $n_L$ with angular momentum $L_v$ at
$R$, is related to total monopole density $n$ by $n_L=f_L n$, with
$f_L$ some function of $R$ and $\theta$. In such case, the result
comes out as:
\begin{eqnarray} \label{new_London}
{\big (} \vec \bigtriangledown \times \vec J_M {\big )} _{\vec R}
&=&  \frac{g n_L (L_v/M r) (2\pi r_L) }{\pi r_L^2} \, \hat{R} \nonumber  \\
&=& \eta \, (c/\lambda_L^2) \, \vec E_{\vec R}
\end{eqnarray}
Here $\lambda_L = ( M c^2 /g^2 n )^{1/2} $ is the London
penetration length. Interestingly enough one arrives at the second
London equation with an modification coefficient
\begin{eqnarray}
\eta=  2 f_L \cos\theta  = 2 f_L \, \frac{  L_{EM}}{
\sqrt{L_{EM}^2+L_v^2}} \approx 2 f_L
\end{eqnarray}

To end this discussion, we emphasize again its main point: motion
of plasma monopoles can function similarly to condensed monopoles,
under the important condition that the field is strong enough,
which specifically means that monopoles' bombarding angular
momentum $L_v$ is much less than the field angular momentum
$L_{EM}$. The effect discussed here and the well-known Meissner
effect of (dual) superconductor share the same mechanism, namely
magnetic particles are scattered by the Lorentz force, inducing
currents which anti-screen the field. On the other hand, there is
a big difference between the two. Meissner effect is only present
in superconductors, which expel arbitrary weak magnetic field.
Plasmas do not have it, and thus distributed weak fields can be
present in the bulk. However, strong enough fields can be expelled
by the plasma into flux tubes, which are in principle metastable.
Solar plasma (for example) has both weak distributed magnetic
fields as well as magnetic flux tubes (visible in telescopes as
substructure of famous solar ``dark spots").

Let us now further simplify the problem, by removing
the electric charges to infinity
 and leaving only a flux of electric field,
confined in a flux tube (see Fig.\ref{fig_demonstration} right).
Again, monopoles from outside which move into it are turned away
by Lorentz force and leave. Although their energy remains
unchanged, the momentum is changed, which means that there is a
constant pressure acting from the monopole ensemble on the tube.
In essence, it is just dual to the Meissner effect of electric
superconductor, in which the magnetic field gets pushed away.

The situation however is only simple if the strength of the field
in the tube is such that particles penetrate only small part of
its radius. (One can then further simplify the problem into flat
surface, with effective current floating in wall separating
field-free and field regions, as is the case for magnetic flux
tube in solar plasma.) As seen from the
Fig.\ref{fig_demonstration} above, if particles can penetrate into
the flux tube beyond its center, they start generating a
counter-rotating current which eventually destroys the tubes. To
further shed light on how the microscopic flux tube may or may not
exist, more careful analysis of the induced current during
scattering of these bombarding monopoles will help.

The lower right picture (transverse projection of the upper) shows
a few trajectories (with different ending points $A,B,B',C,C'$) in
the constant field $E$ region (within tube radius $R$), starting
from the same initial point $I$ (at the bottom) with same velocity
$v$ (thus curled with same Larmor radius $r_L=Mvc/gE$), yet with
different impact parameter $b$ as they are aiming at different
angles. The impact parameter is related to the monopole's incoming
angular momentum by $L_v=\pm Mvb$ depending on whether the
velocity orients toward left or right at the initial point (noting
the positive $\hat z$ is pointing into the page in the figure).
For example, the red curve($I\to A$) is for $b=0$ and $L_v=0$, the
blue solid/dashed curves is for certain nonzero $b$ and same
$|L-v|$ yet opposite signs (with $I \to B$ positive and $I \to B'$
negative), and the green solid/dashed curves for even larger $b$
and $|L_v|$ (with $I \to C$ positive and $I \to C'$ negative).

We now consider the currents $J_b$ produced by various
trajectories. In particular let's examine how the combined current
$j_b$ of two trajectories with same $b$ and $\pm |L_v|$ changes
with $b$. The important observation is the following: trajectories
with {\it small $b$ or small $|L_v|$}, like $I \to A$, contribute
{\it counterclockwise currents}, while trajectories with {\it
large $b$ or large $|L_v|$}, like $I \to C,C'$, contribute {\it
clockwise currents}, and there is a {\it critical $b$ or $|L_v|$
}(which is precisely the solid blue curve $I \to B$ going right
through the center of tube) beyond which the current inversion
happens. Some simple algebra leads to the following {\it critical
angular momentum of the current inversion}:
\begin{equation} \label{critical_L}
|L_c|=mvb= \frac{g \Phi_E}{2\pi c}
\end{equation}
with $\Phi_E=E\pi R^2$ the electric flux. By interpreting the
right-hand-side as the electric-magnetic field angular momentum in
this cylindrical setting, we simply have critical momentum
$|L_c|=L_{EM}$ which coincides with the analysis in the previous
example. Very importantly, strong electric field means large
$|L_c|$ and stable flux tube, while weak field (with small
$|L_c|$) prefers becoming diffusive in the bulk rather than
expelled into flux tube.

As we will show later in the paper, this current inversion
phenomenon is very important. The  counterclockwise currents (from
small $L_v$) strengthen the original field\footnote{Note again
that in the (dual) superconductor case \cite{Ripka}, the Abrikosov
vortex is exactly supported by supercurrent of scattered
condensate in {lowest possible angular momentum, namely $L_v=0$}
channel.}, while the clockwise currents (from large $L_v$) weaken
them. Thus the current inversion is like a "para/dia-electric"
inversion, in macroscopic language, and it kills the flux tubes.

To summarize the lesson from this classical example, the value of
the angular momentum plays essential role in the monopole
scattering by the flux tube. If particles have typical momentum
$\bar{p}$ and the radius of the tube is $R$ the angular momentum
is $\bar{L}\sim \bar{p}R$. When $\bar{L}$ is small or equivalently
the electric field is strong, the motion is still basically radial
and the pressure argument works. In the opposite limit of large
$\bar{L}\sim \bar{p}R >> L_{EM}$ or weak field, the induced
currents have both signs and cancel each other, and there is no
reason for flux tube to exist. Thus there must be some $critical$
value of $\bar{p}R$ above which there is no flux tube solution,
depending on exact magnitude of currents induced in channels with
different angular momentum, to be evaluated quantum mechanically
below.
\\

The rest of the paper is structured as following: by devising and
solving a generalized London's equation, we first show in
Section.II how electric flux tube solution could follow from
"macroscopic" electrodynamics in medium beyond superconductor; in
Section.III we outline our self-consistent treatment of flux tube
starting from "microscopic" level; the quantum mechanic scattering
of single monopole in flux tube field will be exactly solved with
analytic wave functions presented in Section.IV for both
non-relativistic and relativistic cases; we then proceed in
Section.V to impose self-consistent condition and find flux tube
size in a thermal medium; application of our results to sQGP
problem is discussed in Section.VI; and finally Section.VII is for
conclusions.

\section{Electric Flux Tube: Macroscopic approach}

Borrowing wisdom from  electrodynamics of a superconductor and
being motivated by the ``modified London'' relation
(\ref{new_London}) we discussed above, we find in this section
solutions to macroscopic electrodynamics equations of London's
type.

Our generalized (dual) version of the second London equation
reads:
\begin{equation} \label{london_curl}
\vec \bigtriangledown \times  \vec J_M =
\frac{c}{\lambda^{2-\kappa} r^\kappa} \vec E
\end{equation}
Any constant coefficient could be absorbed in a re-definition of
$\lambda$.  When combined with one of the (dual) Maxwell's
equations $\vec \bigtriangledown \times \vec E = -\frac{1}{c} \vec
J_M$, it yields the equation for the electric field
\begin{equation} \label{london_2}
\vec \bigtriangledown^2 \vec E = \frac{1}{\lambda^{2-\kappa}
r^\kappa} \vec E
\end{equation}
Macroscopic parameter $\kappa$ characterizes how the electric
field gets modified by the magnetic medium. $\kappa=0$ is the
London limit (appropriate for the medium being a dual
superconductor in extremely type-I regime), while $\kappa=1$
corresponds to the classical monopole gas (as discussed in
preceding section). Intermediate values of $\kappa$ are suggested
as an interpolation between the two limits, say to describe a
medium having both Bose condensed and non-condensed components.

Our setup corresponds to cylindrical flux tube (see
Fig.\ref{fig_demonstration} right), with $\vec E = E(r) \hat z$ in
coordinates $(r,\phi,z)$. The total electric flux is $\Phi_E=
\int_0^{\infty} E(r) 2\pi r dr $.

The solution for any  $\kappa < 2$  is given by\footnote{For
$\kappa\ge 2$ the boundary condition couldn't be satisfied.}
\begin{eqnarray} \label{flux_tube}
E(r)&&= f_{\kappa} \,\, \frac{\Phi_E}{\pi \lambda^2} \,\,
K_0 {\bigg [}\frac{2}{2-\kappa} (\frac{r}{\lambda})^{(2-\kappa)/2}{\bigg]} \\
f_{\kappa} &&= 1/ {\big[}(2-\kappa)^{\frac{2+\kappa}{2-\kappa}}
\,\, \Gamma[2/(2-\kappa)]^2{\big]} \nonumber
\end{eqnarray}
with $K_0[x],\Gamma[x] $ being the Bessel and Euler Gamma
functions. If a function $F_0[r/\lambda]$ is a solution to London
eq.(\ref{london_2}) with $\kappa=0$, then the function $F_\kappa
[r]\propto F_0{\big [}\frac{2}{2-\kappa}(r/\lambda)^{(2-\kappa)/2}
{\big ]}$ is a solution to the modified eq.(\ref{london_2}) with
any $\kappa$. The normalization constant follows from the total
flux value.

At large distance, $r \to \infty$, the electric field
\begin{equation}
E \sim exp{\big [}-(r/\lambda)^{(1-\kappa/2)}/(1-\kappa/2){\big ]}
{\bigg /} (r/\lambda)^{(2-\kappa)/4}
\end{equation}
vanishes quicker than exponential, leaving most of the flux within
$r \sim \, few \, \lambda$. The smaller is $\kappa$, the thinner
is the flux tube.

For this flux tube solutions the "string tension" -- the energy
per unit length along $\hat{z}$ is
\begin{eqnarray} \label{tube_tension}
\sigma_{\kappa} && = \int_0^{\infty} \frac{E(r)^2}{2} 2\pi r dr =
\frac{\Phi_E^2}{\pi \lambda^2} {\cal T}_{\kappa} \\
{\cal T}_{\kappa} &&= \sqrt{\pi} {\bigg /} {\bigg [}
2^{(6-\kappa)/(2-\kappa)} \cdot (2-\kappa)^{(2+\kappa)/(2-\kappa)}
\cdot \nonumber \\
&& \qquad \qquad \Gamma[(6-\kappa)/(4-2\kappa)] \cdot
\Gamma[2/(2-\kappa)]^2 {\bigg ]} \nonumber
\end{eqnarray}
 ${\cal T}_{\kappa}$
is a rapidly decreasing function of $\kappa$, and in particular
${\cal T}_{\kappa=0} / {\cal T}_{\kappa=1} = 3$ which means that
(for fixed $\lambda$) a quantum condensate expels the electric
flux into a flux tube with the tension three times larger than a
classical monopole gas does.

Let's summarize the physical picture as well as limitations of the
established solution. Any magnetic medium generically expels the
electric field, as monopoles are back-scattered off it, so there
is possibility for flux tube formation. But different media do
this job with distinct efficiencies, leading to flux tube (if
there is any) with different tensions. The electric flux tube
solutions are rather simple: they describe the problem in terms of
two macroscopic properties of magnetic medium, namely $\kappa$ and
$\lambda$. There is however an important limitation. The
macroscopic approach is suitable only if {\it the electric field
strength (or the electric flux here) is large}, as the detailed
analysis in the introductory examples has shown: we repeat that
strong field makes small Larmor radius of monopoles, thus the flux
tube is a "macroscopic" object and scattering of monopoles happens
basically on the surface. If it is not so, the validity of
eqn.(\ref{london_curl},\ref{london_2}) upon which the solution is
based is no longer justified. The intermediate case between
diffusive weak field and macroscopically strong flux tube requires
a microscopic approach, to be discussed in the following sections.\\ \\

\section{The Microscopic Approach}

Starting here and following in subsequent sections, we will pursue
a fully quantum mechanical microscopic approach. Let us first
describe our strategy and approximations made in this section. The
main one is that mutual interaction among monopoles will be
neglected, as it has been argued that magnetic sector of sQGP at
just above $T_c$ is very weakly coupled, see \cite{Liao_ES_mono}
for more details. What's more, if one assumes the monopoles are of
't Hooft-Polyakov type\cite{thooft-polyakov}, the Coulomb
interaction between monopoles may be largely cancelled (and
exactly cancelled in the BPS limit\cite{BPS,Manton} for static
monopoles) by scalar/Higgs exchange. But the Lorentz force from
electric field\footnote{Although the monopoles in sQGP are built
out of non-Abelian fields $A_\mu^a$, each type of monopoles only
interacts with electric field projected into their corresponding
U(1) (see e.g.\cite{Shnir_book,Ripka}), so the Maxwellian field
description still holds.} cannot be cancelled and this is the only
interaction of monopoles relevant to our approach. The single
monopole scattering on a flux tube will be treated quantum
mechanically. Both non-relativistic and relativistic cases will be
analyzed: there are evidences that monopoles in sQGP are
semi-relativistic, e.g. with $M\sim 2T$
\cite{Marco_ES,Claudia_ES}.

In the following sections we will go through the three steps
below: i) first assume existing flux tube of certain size $R$, ii)
then figure out in great details how individual monopole from
medium will be scattered off it and generate some magnetic
current, and iii) finally use the dual Maxwell's equation relating
the electric field and magnetic current to obtain a
self-consistent equation determining the value of $R$ (and thus
string tension $\sigma$) as a function of medium parameters
$T,n,M$. Below we extend the description of strategy a bit more
step by step.\\ \\
 i) For our purpose the electric flux tube with
flux $\Phi_E$ and size $R$ is described in cylindrical coordinate
$(r,\phi,z)$ by the following field:
\begin{equation}
\vec E =  \left\{ \begin{array}{ccc}  E_{I} \hat{z} = \Phi_E/(\pi
R^2) \,\, \hat{z} & , & r\le R \\ 0 & , & r > R\end{array} \right
.
\end{equation}
The corresponding dual vector potential reads:
\begin{equation}
\vec C = C_\phi \, \hat{\phi} = \left\{ \begin{array}{ccc}
\frac{\Phi_E}{2\pi
R} \, \frac{r}{R} \,\, \hat{\phi} & , & r\le R \\
\frac{\Phi_E}{2\pi R} \, \frac{R}{r} \,\, \hat{\phi} & , & r
> R\end{array} \right .
\end{equation}
The string tension is given by
\begin{equation}
\sigma = \Phi_E^2 / (2\pi R^2)
\end{equation}
According to Dirac quantization,  the flux can be normalized via
$(g\Phi_E) /( 4\pi \hbar c) = d/2 $. While the results obtained
below can be used for general $d$, we are particularly interested
in $d=2$ as is true for adjoint monopoles in sQGP\footnote{There
are strong evidences from lattice study of high-$T$ magnetic QCD
which supports the idea that monopoles in QGP have such charges that $d=2$ and
their total numbers scale as $N_c^2-1$, see e.g.\cite{KorthalsAltes:2006gx}}.\\ \\
ii) A monopole moving in such a field is governed by the following
Hamiltonian:\\
in the non-relativistic case
\begin{equation} \label{Hamiltonian}
{\cal H}_{NR} = (\vec p + \frac{g}{c} \vec C )^2 / (2M)
\end{equation}
while in relativistic case it is
\begin{equation} \label{Hamiltonian}
{\cal H}_{R} =\sqrt{ \left ( \vec p + \frac{g}{c} \vec C \right )
^2 \, c^2 + M^2 \, c^4}
\end{equation}
The conserved quantities are\\
1) total energy $\epsilon$; \\
2) longitudinal momentum $p_z$;\\
3) hence one can use longitudinal energy $\epsilon_z=p_z^2/2M$ and
transverse energy $\epsilon_t=\epsilon-\epsilon_z$ to be conserved
separately;\\
4) the angular momentum $L_{z}=r \, (p_\phi +
\frac{g}{c} C_\phi)$.\\
The conservation of both $\epsilon_t$ and $L_z$ implies that the
monopole is rejected back (unless $L_z=0$) when it approaches the
center of the tube, due to generic centrifugal barrier $\sim L_z^2
/(2Mr^2)$ which dominates at small $r$.\\ \\
iii)  The dual Maxwell equation, $\vec \bigtriangledown
\times \vec E = -\frac{1}{c} \vec J_M$, in cylindrical setup being
\begin{equation}
\frac{d \, E(r) }{dr}= \frac{1}{c} J^{\phi}_M
\end{equation}
can be integrated in $r$ 
\begin{equation}
E(r=0) - E(r=R) = - \frac{1}{c}\int_{0}^{R} J^{\phi}_M dr
\end{equation}
The flux tube  may presumably be approximated by a constant $E$
inside certain radius $r<R_-$ and zero outside $r>R_+$, with
smooth interpolation in between. As an approximation in step(i) we
have used step-like electric field, neglecting the difference
between $R_\pm$. The advantage is that monopole motion in such
field can be  calculated (step (ii)). This shouldn't be a serious
issue as we expect $R_+ - R_- << R$. Thus we take $E(r=0)$ as the
constant field strength $E_I=\Phi_E/(\pi R^2)$ within tube and
send $E(r=R)$ to zero, obtaining the equation to be used in later
section:
\begin{equation} \label{field_current}
E_I=\Phi_E/(\pi R^2)=-\frac{1}{c}\int_{0}^{R} J^{\phi}_M dr
\end{equation}
\\

\section{Quantum Mechanical Motion of a single Monopole }

Quantum mechanical motion of single monopole is described by wave
function $\Psi$ which is a scattering solution to ${\cal H} \Psi =
\epsilon \Psi$ with ${\cal H}$ from eq.(\ref{Hamiltonian}). Making
use of conserved quantities, we may decompose the wave function
into $\Psi=f(r) e^{im\phi} e^{i K_z z}$, with energy\footnote{Here
we first deal with non-relativistic case, while in the last
subsection the treatment will be generalized to relativistic case
which turns out to be rather straightforward.}
$\epsilon=\epsilon_z+\epsilon_t=(\hbar K_z)^2/2M + (\hbar k)^2/2M$
and angular momentum $L_z=m\hbar$. Let's first introduce several
parameters involved later in the solution, including:
\be && \nu=m+d\\
&& \gamma=1+|m|\\
&& \alpha=(kR)^2/(4d)-m/2 \ee We repeat that $d= (g\Phi_E)/(2\pi
\hbar c)$ tells how much flux is going through the tube. The
meaning of $\nu$ can be explained as follows: it is quantized
(integer-valued) form of a relation between velocity, canonical
momentum and dual field $m\vec v=\vec p+g\vec C$ projected (via
their cross-product to $\vec r$) to angular momenta. Classical
path $I\to A$ in Fig.1 which has velocity at large distances
directed to the tube center corresponds to $\nu=0$. The $m=0$
channel is the one corresponding to $I\to B$ path: it goes through
the center because it experiences no centrifugal barrier $\sim
m^2/r^2$. As we will see below, this correspondence will explain
the signs of the currents, generated in each partial waves.

The Schrodinger equation can then be reduced to the following
cylindrical radial equation
\begin{equation} \label{radial_eqn}
\frac{1}{r}\frac{d}{dr} \left ( r \frac{d \, f }{dr} \right ) +
\left [ k^2 - V_{eff} \right ] f_{k,\nu} = 0
\end{equation}
The effective potential takes the form:
\begin{equation}
V_{eff} = \frac{1}{r^2} \times \left\{ \begin{array}{ccc}
\left [ \nu+ d \cdot ( r^2/R^2 -1 ) \right ]^2 & , & r\le R \\
\nu^2 & , & r
> R\end{array} \right .
\end{equation}

\begin{figure}[t]
  \epsfig{file=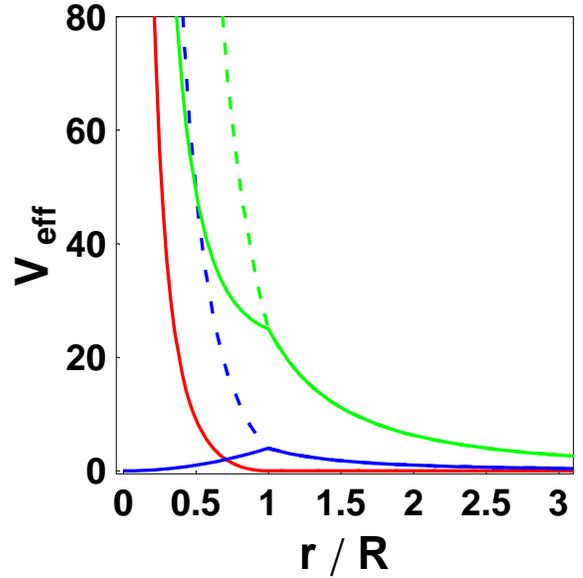,width=7.5cm}
  \vspace{0.15in}
\caption{(color online) The effective potential $V_{eff}$ as a
function of $r/R$ for $\nu=0$(red), $\nu=\pm 2$(blue
solid/dashing), and $\nu=\pm 5$(green solid/dashing). }
\label{fig_veff}
\end{figure}

The equation could be exactly solved both inside and outside the
tube, as shown separately below:\\
i) $r \le R$ (inside), the
solution is given by confluent hypergeometric function $_1F_1[x]$:
\begin{equation} \label{wave_inside}
f^{(i)}_{k,\nu}= A_{k,\nu} \,\, e^{-\frac{d r^2}{2 R^2}} \, \left
( \frac{d r^2}{R^2} \right ) ^{\frac{\gamma-1}{2}} \,  _1F_1 \left
[ \gamma/2-\alpha, \gamma, \frac{d r^2}{ R^2}  \right ]
\end{equation}
ii) $r>R$ (outside), the solution is expressed by two Hankel
functions $H^{(1,2)}_{\nu}[x]$ with proper phase shift
$\delta_{k,\nu}$:
\begin{equation} \label{wave_outside}
f^{(ii)}_{k,\nu}= \frac{B_{k,\nu}}{2} \left [ H^{(2)}_\nu \left[
kr \right] + e^{i 2 \delta_{k,\nu} } H^{(1)}_\nu \left[ kr \right]
\right ]
\end{equation}

Finally the two functions should be connected smoothly at $r=R$,
which
determines: \\
the normalization constants $A,B$ satisfying (with
$J_\nu[x],Y_\nu[x]$ Bessel functions)
\begin{eqnarray} \label{normalization}
{\cal R}_{AB}&&= \frac{A_{k,\nu}}{B_{k,\nu}} \nonumber \\
&&= e^{i \delta_{k,\nu}} \, \frac{  (\cos \delta_{k,\nu}) J_\nu
[kR] - (\sin \delta_{k,\nu}) Y_\nu [kR]}{e^{-d/2} \,
d^{(\gamma-1)/2} \, _1F_1[\gamma/2-\alpha,\gamma,d]}
\end{eqnarray}
and the phase shift $\delta_{k,m}$ being
\begin{eqnarray} \label{phase_shift}
\delta_{k,\nu}&&=\arctan \left[ \frac{J_{\nu+1}[kR] - G
J_\nu[kR]}{Y_{\nu+1}[kR] - G Y_\nu[kR]} \right ] \\
G&&= \left [ \nu - (\gamma-1-d) -(1-2\alpha/\gamma)\cdot d \cdot
\tilde{F} \right ] {\bigg /} (kR) \nonumber \\
\tilde{F}&&= \, _1F_1[\gamma/2-\alpha+1,\gamma+1,d] {\bigg /} \,
_1F_1[\gamma/2-\alpha,\gamma,d]\nonumber
\end{eqnarray}
However exceptions to eq.(\ref{normalization},\ref{phase_shift})
can occur when it so happens that
$_1F_1[\gamma/2-\alpha,\gamma,d]=0$. In such situation the
alternative equations are the following:\\
\begin{eqnarray} \label{normalization_2}
{\cal R}_{AB}&&= \frac{A_{k,\nu}}{B_{k,\nu}}  \\
&&= \frac{ e^{i \delta_{k,\nu}} \, (kR)  \left[ (\cos
\delta_{k,\nu}) J_{\nu+1} [kR] - (\sin \delta_{k,\nu}) Y_{\nu+1}
[kR] \right ] }{e^{-d/2} \, d^{(\gamma+1)/2} \, (2\alpha/\gamma-1)
\, _1F_1[\gamma/2-\alpha+1,\gamma+1,d]} \nonumber
\end{eqnarray}
\begin{eqnarray} \label{phase_shift_2}
\delta_{k,\nu}&&=\arctan \left[ \frac{J_\nu[kR]}{Y_\nu[kR]} \right
] \qquad \qquad \qquad \qquad \qquad \qquad
\end{eqnarray}

The coefficient $B_{k,\nu}$ should be determined by calculating
the current at $r\to \infty$ and matching the physical boundary
current, see more discussions in subsection A. below.\\

To this point, our problem of finding quantum mechanic solutions
(with arbitrary $k,\nu$) for monopole scattering off flux tube
have been all set. With these analytical solutions at hand, a few
discussions are in order below.\\

\subsection{Scattering Amplitude}
Now we discuss the boundary condition and determine the scattering
amplitude. As a scattering problem, we expect an incident current
described by transverse plane wave, say $e^{i k x}$, in the
cylindrical setup. Thus we write down the asymptotic wave function
as\footnote{Here we temporarily normalize the incoming current as
just $v=\hbar k /M$ while in later section additional factor from
density $n$ will be included.}
\begin{equation}
\Psi_k (r \to \infty) = e^{i k x} + \left[\sum_{\nu} {\cal
F}_{k,\nu}(\phi) \right ]  \frac{e^{i k r}}{\sqrt{r}}
\end{equation}
Expanding $e^{i k x}=e^{i k r \cos \phi}$ also in terms of $e^{i m
\phi}$ and comparing the above to the large $r$ limit of
$f^{(ii)}_{k,\nu}(r)$ from eq.(\ref{wave_outside}), we obtain the
normalization constant $B$ as
\begin{equation} \label{coe_b}
B_{k,\nu}= e^{i \pi (\nu / 2 - d )}
\end{equation}
with the feature $|B_{k,\nu}|^2 = 1$ independent of $k,\nu$
values\footnote{One should keep open mind in that different
boundary conditions lead to different weights $B_{k,\nu}$ among
partial waves. It is not clear if there could be choices other
than the ones used here which can best describe the thermal
monopole scattering by flux tube field. An extreme example is
superconductor which picks only $B_{k,0}$ with all others
vanishing.}.

The partial-wave scattering amplitude is determined via phase
shift as
\begin{equation}
{\cal F }_{k,\nu}(\phi)= \frac{e^{-i \pi/4}}{\sqrt{2\pi k}} \,
\left[ e^{i (2\delta_{k,\nu}-d\pi)} -1 \right ] \, e^{i m \phi}
\end{equation}
This gives the partial-wave scattering cross section, or more
precisely transverse cross "length", as
\begin{equation} \label{cross_section}
S_{k,\nu} = \frac{4}{ k} \, \sin^2 (\delta_{k,\nu}-d\pi/2)
\end{equation}
The total cross section is a sum of the above over all $\nu$.

\begin{figure}[t]
  \epsfig{file=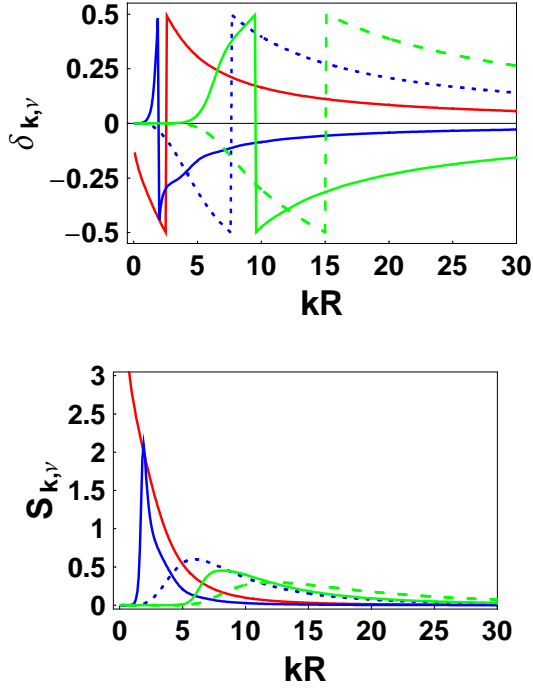,width=7.cm}
  \vspace{0.1in}
\caption{(color online) (upper) Scattering phase shift
$\delta_{k,\nu}$ and (lower) scattering cross section $S_{k,\nu}$
as a function of $kR$ for $\nu=0$(red), $\nu =\pm 2$(blue
solid/dashing), and $\nu=\pm 5$(green solid/dashing).}
\label{fig_scattering}
\end{figure}

Examples of $\delta_{k,\nu}$ and $S_{k,\nu}$ as functions of $k$
for several values of $\nu$ are plotted in
Fig.\ref{fig_scattering}.

Before closing this subsection,we'd like to point out that the
phase of coefficient given in (\ref{coe_b}) is related to the
choice of $e^{ikx}$ as asymptotic incident state (while its unity
amplitude is general). Physically an incident particle can come in
from any direction besides $\hat x$ axis, with equal probability,
so an average over all possible orientation of initial $\vec k$ is
called for. This can be achieved by first doing calculation using
(\ref{coe_b}) and averaging over the $\phi$ dependence at the end,
and the effect of this procedure is simply the entire suppression
of interference terms among different partial waves.\\

\subsection{Magnetic Current}
The magnetic current generated by single monopole during
scattering process can be calculated by
\begin{eqnarray}
\vec j_M &=& \frac{ig\hbar}{2M} \left[ (\vec \bigtriangledown
\Psi^*)\Psi -
 (\vec \bigtriangledown \Psi )\Psi^*  \right ] + \frac{g^2}{Mc} (\Psi^* \Psi)\vec C
\end{eqnarray}
The nontrivial part is the $\hat{\phi}$ component:
\footnote{$\hat{r}$ component is zero and $\hat{z}$ component is
totally irrelevant and can also be set to zero by replacing
$e^{iK_z z}$ with real $\sin,\cos$ functions.}
\begin{eqnarray} \label{mag_current}
J^{\phi}_M (k|r)&&=\sum_{\nu=-\infty}^{\infty} \, j^{\phi}_M (k,\nu|r) \\
&&=\frac{g\hbar}{Mr} \, \sum_{\nu=-\infty}^{\infty} \, \left [ \nu
- d\cdot (1-\frac{r^2}{R^2})\cdot \theta[1-\frac{r}{R}] \right ]
\, |f_{k,\nu}|^2 \nonumber
\end{eqnarray}
with $\theta[x]$ the unit step function. We may further combine
$\pm |\nu|$ terms and rewrite it as
\begin{eqnarray} \label{mag_current_2}
&& J^{\phi}_M (k|r)=\frac{g\hbar}{Mr} \, \sum_{\nu=0}^{\infty} \, \tilde{j}^{\phi}_M (k,\nu|r) \\
&& \tilde{j}^{\phi}_M(k,0|r) = - d\cdot (1-\frac{r^2}{R^2})\cdot
\theta[1-\frac{r}{R}]\, |f_{k,0}|^2 \nonumber \\
&& \tilde{j}^{\phi}_M(k,\nu|r) =  \nu \cdot
\left [ |f_{k,\nu}|^2-|f_{k,-\nu}|^2 \right ] \nonumber \\
&& \qquad \qquad \quad - d\cdot (1-\frac{r^2}{R^2})\cdot
\theta[1-\frac{r}{R}] \cdot \left [ |f_{k,\nu}|^2 + |f_{k,-\nu}|^2
\right ] \nonumber
\end{eqnarray}

This expression implies two important points: first, significant
contribution to magnetic current comes from small $r$ part as is
evident from $1/r$ dependence, so partial waves with large
amplitude at small $r$ (namely small $|\nu|$ channels) are
important; second, according to $j^{\phi}_M \propto \nu$ at $r>R$,
partial waves with $m$ symmetric to $d$, namely a pair of $\pm
|\nu|$ channels, tend to produce opposite currents which
substantially cancel each other. It is worth emphasizing that only
$\nu=0$ partial wave (the one picked by the whole condensate in
ANO vertex case, see e.g. \cite{Ripka}) will benefit from the
first point and at the same time NOT suffer from the second point.

Clearly for each given $k$ the total current $J^\phi_M (r)$ to be
integrated in eq.(\ref{field_current}) should be built up from
summing currents of all partial waves , namely summing $j^\phi_M
(k,\nu|r)$ over quantum numbers $\nu$. One is naturally concerned
with the convergence of such infinite summation, which is
basically determined by the large $|\nu|$ behavior. We can expect
that large $|\nu|$ partial waves contribute very little to the
total current, which bears two simple physical arguments: from
energy point of view, states with $|\nu|$ experience centrifugal
potential $V(r) \sim \hbar ^2 |\nu|^2/2M r^2$ while the kinetic
energy being $E_k = \hbar^2 k^2 / 2M$, so if $kR<|\nu|$ then $E_k
< V(r=R)$ which means it is very hard for the particle to "climb"
up the potential barrier all the way into the tube; from the
impact parameter perspective, states with $|\nu|$ and $k$ have
semiclassical impact parameter $b \sim |\nu| / k$, so if
$kR<|\nu|$ then $b>R$ which means the incident particle will be
largely missing the central part and thus very little scattered,
leading to negligible induced currents. This conclusion has been
confirmed by extensive numerical calculation and practically for
given $kR$ all partial waves with $\nu \ge 1.5 kR$ are vanishingly
small, as is evident from Fig.\ref{fig_j_k_s} to be explained in
next subsection.\\ \\

\subsection{The Total Current}

Now we perform the radial integration
 needed in eq.(\ref{field_current}):
\begin{eqnarray} \label{int_current}
&& \int_0^R J^{\phi}_M dr = \frac{g\hbar}{M} {\cal I}(kR)
=\frac{g\hbar}{M}  \sum_{\nu=0}^{\infty} {\cal I}_\nu (kR) \\
&& {\cal I}_\nu (kR) = \int_0^R dr \,
\tilde{j}^{\phi}_M(k,\nu|r)/r \nonumber
\end{eqnarray}

\begin{figure}[t]
  \epsfig{file=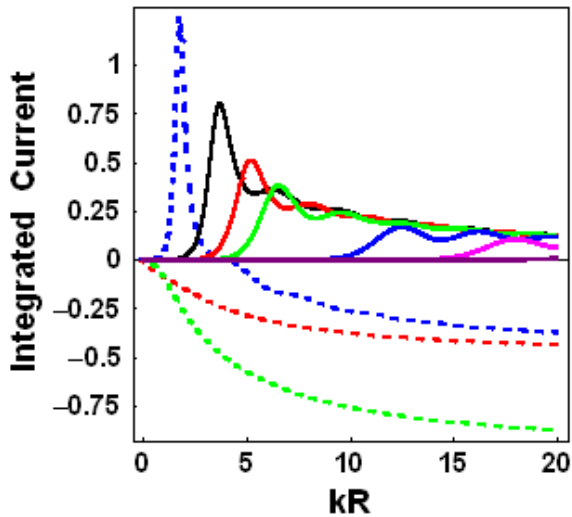,width=7.5cm}
  \vspace{0.1in}
\caption{(color online) Integrated current ${\cal I}_\nu$ as a
function of $kR$ for different values of $\nu$: dashed lines are
for $\nu=0$(red), $\nu=1$(green), $\nu=2$(blue), while solid lines
are for $\nu=3$(black), $\nu=4$(red), $\nu=5$(green),
$\nu=10$(blue), $\nu=20$(magenta), and $\nu=30$(purple). }
\label{fig_j_k_s}
\end{figure}

In Fig.\ref{fig_j_k_s} we plot ${\cal I}_\nu$ versus $kR$ for
various $\nu$. The interesting observation is that the integrated
current is negative for $\nu<2$, positive for $\nu>2$, while for
$\nu=2$ partially positive (at small $kR$) and negative (at large
$kR$). This result from quantum mechanics perfectly agrees with
our conclusion from classical treatment in the Introduction part,
not only qualitatively but even quantitatively: the critical
angular momentum for current inversion observed here $L_c= \nu_c
\hbar$ with $\nu_c=2$ coincides with that predicted by
eq.(\ref{critical_L}) once our flux $g\Phi_E=d \, 2\pi \hbar c$
with $d=2$ is plugged in.\\ \\

\begin{figure}[t]
  \epsfig{file=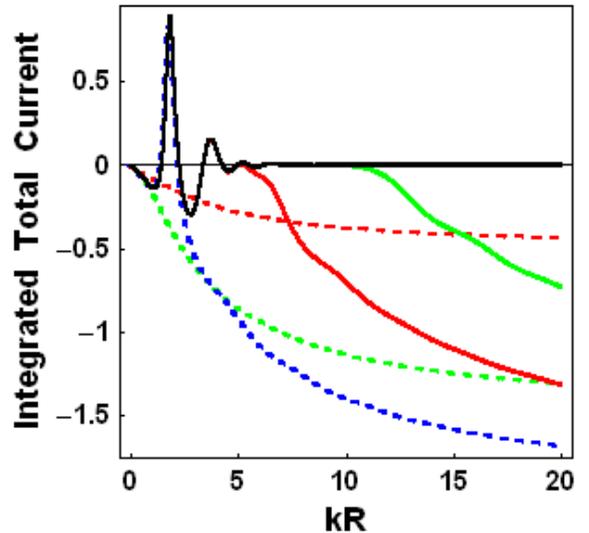,width=7.5cm}
  \vspace{0.1in}
\caption{(color online) Integrated total current ${\cal I}$ as a
function of $kR$ for different values of summation cut
$\nu_{cut}$: dashed lines are for $\nu_{cut}=0$(red),
$\nu_{cut}=1$(green), $\nu_{cut}=2$(blue), while solid lines are
for $\nu_{cut}=5$(red), $\nu_{cut}=10$(green),
$\nu_{cut}=20$(blue), and $\nu_{cut}=30$(black).  }
\label{fig_j_k_t}
\end{figure}

Now we perform the final step: namely summing ${\cal I}_\nu$ over
$\nu$ to obtain the integrated total current ${\cal I}$. This is
done numerically, with summation cut $\nu \le \nu_{cut}$ applied,
see Fig.\ref{fig_j_k_t} for results for various $\nu_{cut}$. As
can be seen, for the displayed regime $kR\le 20$ the summation is
converged enough as soon as $\nu_{cut}\ge 20$, as the curves with
$\nu_{cut}=20 $ and $\nu_{cut}=30$ coincide on top of each other
and are hardly distinguishable. It is this numerically evaluated
function ${\cal I}(kR)$(with our highest cut $\nu_{cut}=30$) that
will be used in subsequent sections.

The behavior of this function ${\cal I}(kR)$ has rather nontrivial
wiggle structure: the general trend is oscillatory, with a modest
negative part at small $kR<1.42$ (basically from negative
contribution from $\nu=0,1$) followed by a rather high positive
peak (dominantly from $\nu=2$) between $1.42 \to 2.24$. These
first two structures, first negative then positive, basically
cover the interesting region of $kR$(see discussion in next
paragraph). Suppose there is a flux tube with certain $R$, then at
low temperature the typical $\bar{k}$ is small and $\bar{k}R$
falls within negative region which supports the flux tube, while
at high temperature the larger $\bar{k}$ brings $\bar{k}R$ beyond
the negative region into the tremendous positive region which will
kill the flux tube. So there is a transition with the border at
$kR=1.42$: beyond this point higher partial waves with $\nu \ge
1.5\times 1.42 =2.13$ (which is also close to the classical
critical value $\nu_c=2$) will become dominant. By Comparison of
this curve with the red dashed one (only $\nu=0$), which
corresponds to what superconductor can do, one understands why a
condensate does much better in confining a flux tube than a normal
thermal ensemble can do. The first negative peak, actually the
best point\footnote{One might argue that there will be an even
larger negative peak at $kR\approx 2.77$, however to reach that
point one requires much larger $\bar{k}$ which usually means
broader distribution over $k$ around $\bar{k}$, and that will
easily make the total contribution rather small after cancellation
with the adjacent large positive peak.} for flux tube formation,
locates at
\begin{equation} \label{peak}
k_m R = 1.076 \quad with \quad {\cal I}(k_m R) = -0.140
\end{equation}

Finally let's discuss interesting range of $kR$. Remember
ultimately we'd like to discuss the flux tube inside an ensemble
of monopoles with temperature $T$. So first, the tube radius
shouldn't be much larger than $\hbar c / k_B T$, otherwise the
tube's transverse vibrational modes ($\omega \sim 1/R$) get too
easily thermally excited, making it unstable. Second, large $k$
should be suppressed by thermal distribution, and typical
$\bar{k}$ should be few times $k_B T /\hbar c$. Thus it follows
that typical values of $\bar{k}R$ should be of the order unity.
The evaluated ${\cal I}(kR)$ up to $kR=20$ here should be
sufficient for later application.

\subsection{Partial Wave with $\nu=d$ and Possible Resonance}

\begin{figure}[t]
  \epsfig{file=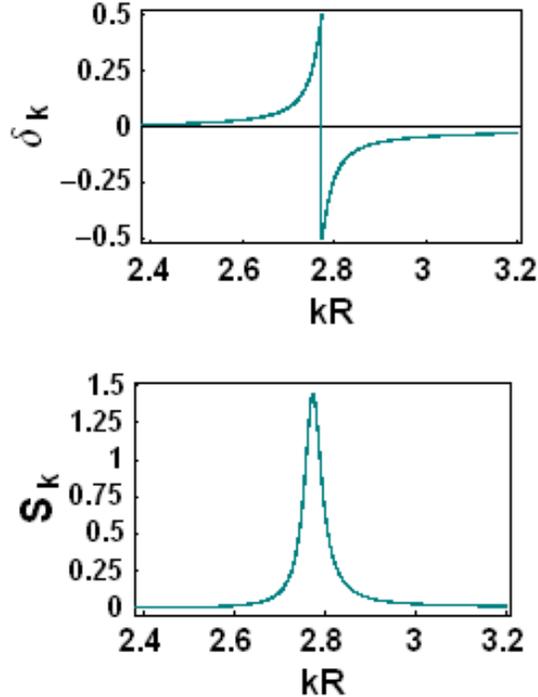,width=7.cm}
  \vspace{0.1in}
\caption{(color online) (Upper) Phase shift and (lower) scattering
cross section as functions of $kR$ for $\nu=d=4$ which show
resonance structure, see text. } \label{fig_resonance}
\end{figure}

The effective potential $V_{eff}$ with $\nu=d$ or equivalently
$m=0$ is special in that it vanishes at the center $r=0$, while
for all other $\nu \ne d$ states there will be diverging term
$\sim (\nu-d)^2 \hbar^2 /2M r^2$. The shape of it (see
Fig.\ref{fig_veff}) actually indicates possibility for resonance
to occur. Whether there could be resonance solution or not depends
on the competition of the localization energy and the potential
barrier whose peak value is $E_{p.}=d^2 \hbar^2/2MR^2$ at $r=R$.
To settle this one can look at the condition for the wave function
(\ref{wave_inside}) to be zero right upon $r=R$ (which is very
close to the resonance situation and gives estimate of kinetic
energy). This yields the series of particular values of $k$: $k_0
R=2.576$, $k_1 R=5.632$, $k_2 R=8.729$, $k_3 R=11.847$, ... Thus
clearly to have one resonance level, one needs at least
$E_{p.}>\hbar^2 k_0^2/2M$ namely $d>2.576$. Indeed by fine search
for resonance structure in scattering phase shift we identified
one resonance in the case of $d=4$, see Fig.\ref{fig_resonance},
with $k_{res.}\approx 2.77$ very close to the above $k_0$ and
narrow width $\Gamma_k \sim 0.1/R$. Nothing similar was found in
$d=2$. With large enough $d$ the occurrence of resonance should be
a general phenomenon and the induced current produced by these
resonance states actually will spoil the original flux tube field
as monopole in such state stays in the center of tube and "pushes"
field outward rather than inward: remember the large positive peak
in ${\cal I}(kR)$ (black curve in Fig.\ref{fig_j_k_t}) is
precisely due to the contribution from $\nu=d$ partial waves.

\subsection{Quantum Mechanical Motion of Single Relativistic
Monopole }

In this subsection we generalize the obtained solutions to
relativistic case. Now one has to solve Klein-Gordon equation
(since monopoles are scalar particles) instead of Schrodinger
equation:
\begin{equation}
\left [ \epsilon^2 - M^2 \, c^4 -  \left ( \vec p + \frac{g}{c}
\vec C \right ) ^2 \, c^2 \right ] \Psi = 0
\end{equation}
Fortunately it turns out that by again writing eigenstate of
energy $\epsilon$ as $\Psi=f_{k,\nu}(r) e^{im\phi} e^{i K_z z}$
one recovers exactly the same radial equation as
eq.(\ref{radial_eqn}) except for changing
$k=\sqrt{(2M\epsilon)/\hbar^2-K_z^2}$ to the following
\begin{equation}
k=\sqrt{(\epsilon^2-M^2 \, c^4)/(\hbar c)^2 - K_z^2}
\end{equation}
So all the exact wave functions obtained in non-relativistic case
are still solutions to the Klein-Gordon equation after the above
replacement of $k$ in eq.(\ref{wave_inside})(\ref{wave_outside}).
This change of $k$ should be done for all the relevant formulae
above.

Another important change is for the current equation
(\ref{mag_current}): due to relativistic effect the mass $M$
should be replaced by $\epsilon/c^2$ , namely
\begin{equation} \label{mag_current_r}
j^{\phi}_M (k,\nu|r)=\frac{g\hbar}{(\epsilon/c^2)r} \, \left [ \nu
- d\cdot (1-\frac{r^2}{R^2})\cdot \theta[1-\frac{r}{R}] \right ]
\, |f_{k,\nu}|^2
\end{equation}
The same replacement should also be applied to integration over
current in eq.(\ref{int_current}).

All other aspects remain pretty much the same as in
non-relativistic case and we skip further discussion.

\section{Self-Consistent Electric Flux Tube Solution}

In this section we will self-consistently determine the size $R$
of the electric flux tube carrying flux $\Phi_E = d \times (2\pi \hbar
c) / g$ in an ensemble of monopoles with temperature $T$ and
density $n$, by using the $integrated$\footnote{Since we are not
  interested in details of the flux tube shape,
we refrain from doing more complicated local matching of the
current and the $\vec \bigtriangledown \times \vec E$, as local
form of Maxwell equation demands. }  magnetic current obtained in
previous section.

At this stage the issue is to average the integrated total current
${\cal I} (kR) $ over proper thermal distribution $n(k|T)$
(through which the medium property comes into play) satisfying
$n=\int_0^\infty dk \, n(k|T)$. We have
\begin{equation} \label{current_fraction}
\frac{1}{c} \int_0^R J^\phi_M dr = \frac{g \hbar}{Mc}
\int_0^\infty dk \, n(k|T) \, {\cal I} (kR)
\end{equation}
Note in relativistic case we have to replace the mass $M$ by
$\epsilon(k)/c^2$ and move it inside the integration over k. Below
we deal with non-relativistic gas, relativistic gas, and optimally
correlated ensemble separately.

\subsection{Non-Relativistic Gas}

In non-relativistic(NR) gas with $Mc^2/k_BT$ large, the kinetics
are simplified, yet in principle one still needs to take into
account the quantum statistics, namely using the Bose-Einstein(BE)
distributions. Only in the non-degenerate limit (with monopole gas
being not dense) one recovers the Boltzmann limit. So we use the
BE distribution $1/(z^{-1}e^\epsilon -1)$ normalized to density
$n$ by
\begin{equation}
n = s \times \left ( \frac{Mk_B T}{2\pi \hbar^2} \right )^{3/2}
\times Li_{\frac{3}{2}}[z]
\end{equation}
In the above $s$ is the degeneracy due to internal degrees of
freedom, fugacity $z=e^{\mu/k_B T}$ is related to chemical
potential and valued as $0\le z < 1 $ in NR case, and
$Li_{3/2}[z]$ is the polylogarithm function. We then have the
$n(k|T)$ given by (after integrating out the $\hat{z}$ momentum)
\begin{equation} \label{nr_dis}
n(k|T)dk = s \times \left ( \frac{Mk_B T}{2\pi \hbar^2} \right
)^{3/2} \times Li_{\frac{1}{2}}\left [z\, e^{-y^2} \right ] \, \,
2y \, dy
\end{equation}
with the variable $y=\hbar k / (2M k_B T)$.

Now by combining
eq.(\ref{field_current})(\ref{current_fraction})(\ref{nr_dis}) we
obtain the self-consistent equation for flux tube size $R$:
\begin{eqnarray} \label{tube_NR_BE_1}
&&E_I=\frac{\Phi_E}{\pi R^2}= \frac{g\hbar n}{M c} \times
\frac{\hbar^2}{ R^2 Mk_BT} \times {\cal U}\left[
q=\frac{R}{\sqrt{\hbar^2/\pi Mk_BT}} \right ] \nonumber \\
&&
\end{eqnarray}
with the last term ${\cal U}$ from integration over $x=kR$
\begin{eqnarray}  \label{tube_NR_BE_2}
{\cal U}[q]= && - \int_0^\infty \, dx \,\, x\,
\frac{Li_{\frac{1}{2}}[ z \, e^{-\pi x^2 /
2q^2]}}{Li_{\frac{3}{2}}[z]} \, \,  {\cal I}(x)
\end{eqnarray}

The self-consistent equation can be further rewritten in an
elegant way:
\begin{equation} \label{self_con_nr}
 \frac{2d}{\pi}  \times
\left(\frac{\lambda_L}{\lambda_{dB}}\right)^2 = {\cal U}\left [
q=\frac{R}{\lambda_{dB}}\right ]
\end{equation}
with $\lambda_L=(Mc^2/g^2 n)^{1/2}$ and $\lambda_{dB}=(\hbar^2/\pi
M k_B T)^{1/2}$. So for given parameters one uniquely determines
the flux tube size $R$ from the above equation.

The NR Boltzmann limit, satisfying scale hierarchy $1/n^{1/3} >>
\hbar /\sqrt{Mk_B T}
>> \hbar / (Mc) $, can be achieved by simply replace $\frac{Li_{\frac{1}{2}}[ z \, e^{-\pi x^2 /
2q^2]}}{Li_{\frac{3}{2}}[z]}$ in the integration of ${\cal U}[q]$
by $e^{-\pi x^2 / 2q^2}$. Mathematically this follows from taking
the $z\to 0$ limit (with only linear terms left) of both
polylogarithm functions.

\begin{figure}[t]
  \epsfig{file=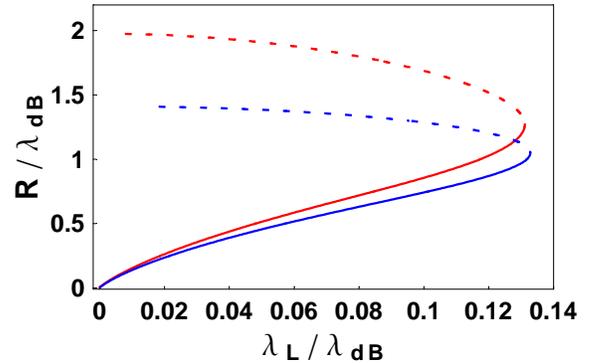,width=7.5cm}
  \vspace{0.1in}
\caption{(color online) $R/\lambda_{dB}$ versus
$\lambda_L/\lambda_{dB}$ from solution of eq.(\ref{self_con_nr}).
The blue curve is for Boltzmann limit ($z\to 0$) while the red for
$z=0.999$, see text. } \label{fig_rlambda}
\end{figure}

The results from solving eq.(\ref{self_con_nr}) are plotted in
Fig.\ref{fig_rlambda}. Numerically we didn't see much difference
between $z \to 0$(blue curve) and $z \to 1$(red curve) limits. As
$\lambda_L/\lambda_{dB} \propto (T/n)^{1/2}$, the right end of the
horizontal axis corresponds to high-density/low-temperature regime
while the left end represents low-density/high-temperature
regime.\\

The distinguished feature is the existence of critical point for
$\lambda_L/\lambda_{dB}$ beyond which there will be no
self-consistent solution: this occurs at roughly the same value
for both displayed curves and we obtain the following condition
for the existence of flux tube
\begin{equation} \label{critical_NR}
\frac{\lambda_L}{\lambda_{dB}} =\left ( \frac{\pi M^2 c^2 k_B
T}{\hbar^2 g^2 n} \right )^{1/2} \le 0.13
\end{equation}

Physically the above result is very appealing: it demonstrates the
mechanism of how a flux tube which exists in the medium at low $T$
is eventually gone as the medium is heated up; on the other hand,
for a medium with given $T$ it sets up a lower bound of monopole
density that is required to support the existence of flux tube.

Another feature is that for each given $\lambda_L/\lambda_{dB}$
smaller than the critical value, there are actually two solutions,
one with small $R/\lambda_{dB}$ (typically smaller than 1, see
solid curves) and the other with large $R/\lambda_{dB}$ (typically
greater than 1, see dashed curves). This is understandable
according to the complicated wiggle structure of ${\cal I}(kR)$.
The solution with smaller radius is the stable one: it is much
thinner and thus has stronger electric field ($E\sim \Phi_E /
r^2$), which reflects monopoles more sharply near the boundary.
The other solution with larger $R/\lambda_{dB}$ is unstable and
should be discarded.

\subsection{Relativistic Gas}
In relativistic gas the important scale is set by temperature, so
let's introduce the following dimensionless variables:
\begin{equation}
w=\frac{R}{\hbar c/k_B T} \quad ,  \quad u=\frac{Mc^2}{k_B T}
\quad
\end{equation}
The fugacity $z$, now in range $0<z<e^{u}$, is related to density
by
\begin{eqnarray}
n &&= {\it s} \times \left ( \frac{k_B T}{\hbar c} \right ) ^3
\int_0^\infty \frac{t^2 \, d t / (2 \pi^2 )}{z^{-1} e^{\sqrt{u^2+
t^2}}-1} \nonumber \\ &&= {\it s} \times \left ( \frac{k_B
T}{\hbar c} \right ) ^3 \times {\cal C}
\end{eqnarray}
with the number ${\it s}$ the degeneracy due to internal degrees
of freedom. ${\cal C}$ serves as normalization constant to
momentum distribution (after scaling momenta by $k_B T /c$).

The distribution over $k$ is given by
\begin{equation}
n(k|T) dk dK_z= {\it s} \times \frac{ k \, dk \, dK_z / (4
\pi^2)}{z^{-1} e^{\epsilon(k,K_z)/k_B T}-1}
\end{equation}
with $\epsilon(k,K_z)=\sqrt{M^2c^4+\hbar^2 k^2 c^2+\hbar^2 K_z^2
c^2}$. Similarly combining the above with
eq.(\ref{field_current})(\ref{current_fraction}) one obtains the
relativistic version of the self-consistent equation
\begin{eqnarray} \label{self_con_r}
&&E_I=\frac{\Phi_E}{\pi R^2}= \frac{g \, n   }{4\pi^2 \, (k_B T /
\hbar c)} \times {\cal U}[w ]
\end{eqnarray}
with ${\cal U}[w]$ given by the following integral
\begin{eqnarray}
{\cal U} [w] =&& - \frac{1}{{\cal C}} \int_0^{\infty} dx \, x \,
{\cal I} (w \,
x) \nonumber \\
&& \qquad \qquad \times \int_{-\infty}^\infty dy \,
\frac{1/\sqrt{u^2+x^2+y^2}}{z^{-1} e^{\sqrt{u^2+x^2+y^2}}-1}
\end{eqnarray}
We can further rewrite the self-consistent equation as
\begin{equation}
8\pi^2 d  \times \left (
\frac{\tilde{\lambda}_L}{\tilde{\lambda}_{dB}} \right )^2 = w^2 \,
{\cal U}[w|u,z]
\end{equation}
with the newly introduced relativistic parameters
$\tilde{\lambda}_L=(k_B T / g^2 n)^{1/2}$ and
$\tilde{\lambda}_{dB}=\hbar c / k_B T$.

For given sets of parameters $M,n,T$ (or equivalently $u,z,T$) one
can easily find the flux tube size $R$ from the above equations by
direct numerics. The situation is quite similar to the
non-relativistic gas which we skip further discussion.

\subsection{Optimally Correlated Ensemble}

Finally let's discuss ensemble beyond an ideal gas. Clearly with
significant interparticle correlations the ensemble may even not
be easily describable by any distribution, however a typical
momentum $\bar{k}_T$ can still be invoked. A special situation
which we call optimally correlated ensemble is that monopoles from
such ensemble are largely carrying momenta within very narrow
region around $\bar{k}_T$. On the contrary if the ensemble
particles' momenta are very diffusive in momentum space, it can
hardly support flux tube.

In the optimally correlated ensemble, we approximate
eq.(\ref{current_fraction}) as (assuming NR formulae)
\begin{equation} \label{current_correlated}
\frac{1}{c} \int_0^R J^\phi_M dr = \frac{g n \hbar}{Mc} {\cal I}
(\bar{k}_T R)
\end{equation}
and the self-consistent equation is then given by
\begin{equation} \label{self_correlated}
E_I=\frac{\Phi_E}{\pi R^2}= \frac{g n \hbar}{M c} \times [-{\cal
I}(\bar{k}_T R)]
\end{equation}
We limit the value of $\bar{k}_T R$ within $0-1.42$ beyond which
there won't be flux tube solution, as discussed in Section.IV C.

The above can be re-organized into
\begin{equation} \label{self_correlated_2}
2d \times (\bar{k}_T \cdot \lambda_L)^2 = (\bar{k}_T R)^2 \times
[-{\cal I}(\bar{k}_T R)]
\end{equation}
The best situation occurs (roughly) around the negative peak in
${\cal I}(kR)$ given by (\ref{peak}). From this we set a bound
similar to eq.(\ref{critical_NR})
\begin{equation} \label{critical_correlated}
\bar{k}_T \cdot \lambda_L = \left ( \frac{Mc^2 \bar{k}_T^2}{g^2 n}
\right )^{1/2} \le 0.20
\end{equation}

\section{Disappearance of Flux Tubes in  Quark-Gluon Plasma}

Results from previous sections are general in nature and
applicable to a variety of plasma physics problems. The present
section, on the other hand, is dedicated to our main application,
the physics of sQGP. From now on we switch to natural units and
systematically put $\hbar,c,k_B=1$.

\begin{figure}[t]
  \epsfig{file=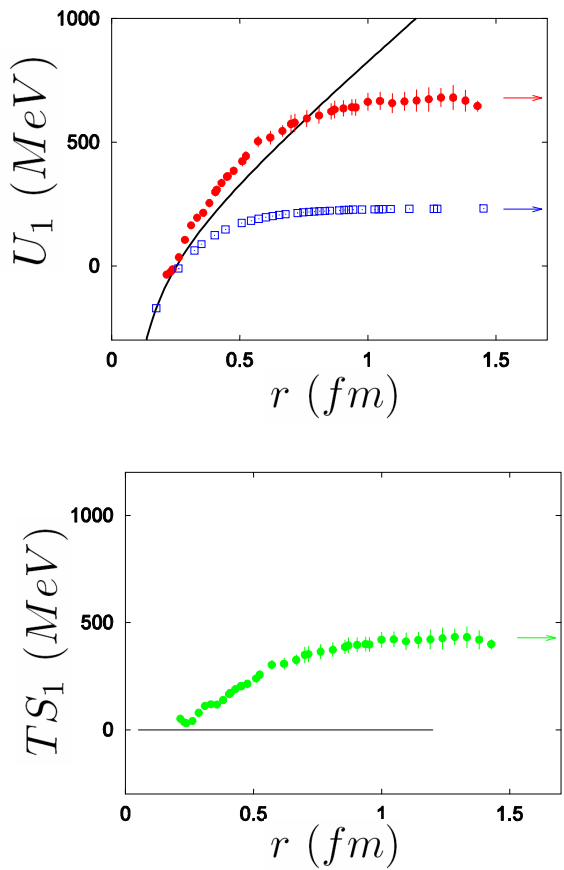,width=7.5cm}
    \vspace{0.1in}
\caption{
  (a) The singlet internal energy, $U_1(r,T)$ (filled circles),
  calculated from renormalized singlet free energy, $F_1(r,T)$ (open squares),
  at fixed $T\simeq1.3T_c$ in $2$-flavor lattice QCD compared
  to $V(r,T=0)$ (line) .
  (b) The corresponding color singlet quark anti-quark entropy,
  $TS_1(r,T\simeq1.3T_c)$, as function of distance calculated from renormalized
  free energies.}
\label{pot1fig}
\end{figure}

The existence of string/flux tubes in the QCD confined phase
$T<T_c$ is rather thoroughly investigated on lattice, via
measurements of static heavy quark potentials. Static free energy
potentials $F(T,r)$ as a function of $r$ are only studied for
$N_c=3$ but for number of quark flavors $N_f=0,2$ as well as
physical QCD, see \cite{lattice_free_energy}. Those can be used to
extract the entropy and potential energy separately: the peaks of
these quantities (see e.g. Fig.2 of \cite{Kaczmarek:2005zp})
happen to be exactly at $T=T_c$ and then decrease toward larger
$T$. The presence of the quasi-linear part of the energy and
entropy at intermediate $r$ leads to a conclusion that flux tubes
still exist at $T>T_c$. Fig.\ref{pot1fig} from
\cite{Kaczmarek:2005zp} shows how both the internal energy and
entropy look like at $T= 1.3 T_c$. Unlike in the free energy (open
squares in the upper plot, in which cancellation takes place), the
internal energy (closed circles) still show at intermediate
$r=(.3-.7) \, fm$ a part linearly dependent on $r$, while at
$T>1.3T_c$ it very quickly disappears.\\

Why are flux tubes disappearing at large $T$? It can in principle
be due to two different changes in QGP happening as T grows above
$T_c$, to be called (i) electric screening and (ii) magnetic
penetration. The simplest mechanism (i) is that as $T$ grows
beyond $T_c$, the density of electrically charged quasiparticles
-- gluons and quarks -- is growing and eventually it becomes large
enough to screen heavy quarks. The reason for this density growth
is the decrease in effective masses of electric excitations, which
are lattice observables by themselves\footnote{See a related
discussion of various color-electric objects' effective masses in
\cite{Liao_ES_susceptibilities}. There we showed the masses are
still rather large and their densities rather small at 1-1.5$T_c$,
limiting the screening.}. At very large $T>>T_c$, in weak
(electric) coupling domain, the screening of the potential is
expected to be described by the Debye theory. However Debye theory
does not describe entropy and internal energy associated with
static quarks at $T=(1-1.3)T_c$, even at large distances, as can
be seen e.g. from calculations of Antonov et al
\cite{Antonov:2006wz}.

Another effect (ii), discussed for the first time in this work,
is the penetration of
 magnetically charged quasiparticles (MQPs) inside the flux tubes,
which destroys them. Indeed, the key parameter $\bar{k} R$
increases with $T$ and reaches the critical value
eq.(\ref{critical_correlated}) for whether flux tube can exist or
not. This imposes the following condition
\begin{equation} \label{critical_sQGP}
\frac{g^2}{4\pi}  (\frac{n}{T^3}) \ge 2.0  \left (
\frac{\bar{k}_T}{T} \right )^2   \frac{M}{T}
\end{equation}
Changing $T$ from $T_c$ upward the monopoles gets heavier and
their dimensionless magnetic density $n/T^3$ keeps decreasing:
eventually this will violate the flux tube condition.
 We thus identify the equality in (\ref{critical_sQGP}) with the
temperature $T \approx 1.3T_c$ at which local dissolution of the
flux tubes takes place.

Furthermore, at $T \approx 1.3 T_c$ we expect $g^2/4\pi \approx 1$
\cite{Liao_ES_mono}. An independent consideration fixes conditions
for monopole Bose condensation \cite{Marco_ES} which demands that
around $T_c$ the monopole mass over temperature $M/T \approx 1\sim
1.2$.

Combining these estimates with our critical condition for tube
dissolution we obtain the density of magnetic quasiparticles at
$1.3 T_c$ to be
\begin{equation} \label{magnetic_density}
n_{MQPs} \approx\left (\frac{n}{T^3} \right ) {\bigg |} _{T=1.3T_c} \approx \quad
2 \sim 3
\end{equation}
which is within $n_{MQPs} = (4.4 - 6.6) fm^{-3}$ in absolute units.

Can the density of magnetic objects really be of that magnitude
(which superficially looks rather high)? This estimated density
includes in principle contributions from all types of magnetically
charged objects in sQGP, i.e. not only pure adjoint monopoles but
also self-dual dyons and also dyons containing quarks\footnote{We
recall that monopoles have fermionic zero modes and states made of
fermions travelling on top of a monopole have to be included as
well. In supersymmetric theories those form spin-1/2 and even 1
magnetic objects, which are needed by supersymmetry to produced
appropriate supermultiplets including the usual scalar
monopoles.}.

Let us compare the numbers with whatever is mentioned in
literature. We don't know any studies of fermionic objects
mentioned above.

Ilgenfritz et al \cite{Ilgenfritz_dyon} determined their dyon
estimate by the caloron density, which is reliably calculated from
the topological susceptibility.  After multiplying by 3/2 their
result for SU(2) we obtain density of self-dual dyons to be
$n_{dyons}\sim 3 fm^{-3}$. Chernodub and Zakharov
\cite{Chernodub_Zakharov} mentioned the monopole density which is
directly estimated from lattice configurations by following
gauge-fixed monopoles along their trajectories. Their estimate is
about $n_{mono}\approx 3.5 fm^{-3}$. The sum of the two is
consistent with the upper end of our estimate of what is needed
for formation/dissolution of the flux tube.

Independent comparison can also be made with the vacuum ($T=0$)
monopole density. Bali \cite{Balistrings} has measured London
penetration length by fitting lattice result with Abelian Higgs
model. From that one can infer the monopole density to be as large
as $10 fm^{-3}$. Bornyakov et al \cite{Bornyakov} gave the vacuum
monopole density to be about $7.5 fm^{-3}$. All these results are
well above our estimates for the density at $T=1.3T_c$
``dissolution point''.

We believe all these numbers are consistent and suggest a coherent
picture, of very dense monopole condensate in vacuum, tightly
confining electric flux into very narrow tubes. When heated
slightly above $T_c$ the monopole condensate changes into a
non-condensed ensemble of monopoles, which is roughly twice less
dense. Yet it is still capable of supporting flux tubes survived
from vacuum, and only around $T=1.3T_c$ the density of monopoles
drops so low that there won't be flux tube any more. At higher $T$
the electric sector becomes more and more dominant till eventually
small number of heavy monopoles become embedded in the
perturbative electric plasma.

\section{Summary and outlook}

In this paper we have  studied stability of the electric flux
tube in a monopole plasma. Quantum scattering of a single monopole on
electric flux tube is analyzed in great details. Already classical
analysis hints on the existence of a critical angular
momentum dividing the scattered magnetic currents which
support/dissolve the flux tube. This finding is quantitatively
confirmed by quantum mechanic calculation, in which we have found
exact scattering solutions to Schrodinger/Klein-Gordon equation in
non-relativisitic/relativistic situations. These solutions allowed us
to calculate  the magnetic current produced, which is
 then averaged over the monopole ensemble and used in
 self-consistent determination of the flux
tube size. The exact critical condition has been established, and
applied to electric flux tube dissolution in sQGP system which
interests us most. This leads to an estimate of total density of
magnetic quasiparticles
 $n_{MQPs}\approx 4.4 \sim
6.6 fm^{-3}$ at $T \approx 1.3 T_c$, where
lattice potentials indicate flux tube dissolution. These numbers
are consistent with
other studies using alternative ways to estimate magnetic density.

 As mentioned in the introduction, this work is partly methodical
in nature, ignoring electric quasiparticles which would lead to
screening and termination of flux tubes. The next step
we plan to do is obviously inclusion of both components
and calculation of the static potentials.
 Hopefully, when one would consider an appropriate mixture of electric and magnetic quasiparticles, the
lattice data on static potentials between electric and
magnetic\footnote{Those are given by the expectation value of the
so called 't Hooft loop: we have not discussed them in this work.}
charges would be explained.

In principle, one should go beyond that and calculate field
distributions around static charges as well. Lattice studies can
be extended to measure directly electric/magnetic fields at $T\sim
T_c$: in fact the field profiles have  been measured  for flux
tubes in vacuum before (see e.g. \cite{Balistrings}).

Let us end  with the following intriguing question. We focused
above on electric flux in magnetic media, ignoring electric
quasiparticles and possible dual phenomenon --
 a magnetic flux tube in an electric plasma. (We only
mentioned their existence at low $T$ high density regime, in a
color superconductor.) Now, may somewhere along the
electric-magnetic equilibrium line there be conditions supporting
stable flux tubes of $both$ types at the same time? It is known
that confinement of both is impossible, but in a uncondensed
plasma regime it may still be the case. A natural place to look
for a QGP with intertwined electric and magnetic flux tubes is at
$T$ less or of the order of $T_c$, close to the place where three
major phases --  hadronic, color superconductor and QGP -- meet.
Although it is quite challenging task to get into this region
using lattice gauge methods, the task is not hopeless.

\vskip .25cm
{\bf Acknowledgments.}
\vskip .2cm

This work was supported in parts by the US-DOE grant
DE-FG-88ER40388.

\end{document}